# Jigsaw-based Secure Data Transfer over Computer Networks

Rangarajan A. Vasudevan
Department of Computer Science and Engineering
Indian Institute of Technology
Chennai 600036 India
Email: ranga@cs.iitm.ernet.in

Sugata Sanyal
School of Technology and Computer Science
Tata Institute of Fundamental Research
Mumbai 400005 India
Email: sanyal@tifr.res.in

Ajith Abraham
Computer Science Department
Oklahoma State University
OK 74106 USA
Email: ajith.abraham@ieee.org

Dharma P. Agrawal
Department of ECECS
University of Cincinnati
OH 45221-0030 USA
Email: dpa@ececs.uc.edu

*Abstract*— In this paper, we present a novel encryption-less algorithm to enhance security in transmission of data in networks. The algorithm uses an intuitively simple idea of a "jigsaw puzzle" to break the transformed data into multiple parts where these parts form the pieces of the puzzle. Then these parts are packaged into packets and sent to the receiver. A secure and efficient mechanism is provided to convey the information that is necessary for obtaining the original data at the receiver-end from its parts in the packets, that is, for solving the "jigsaw puzzle". The algorithm is designed to provide information-theoretic (that is, unconditional) security by the use of a one-time pad like scheme so that no intermediate or unintended node can obtain the entire data. An authentication code is also used to ensure authenticity of every packet.

## I. INTRODUCTION

Security of network communications is arguably the most important issue in the world today given the vast amount of valuable information that is passed around in various networks. Information pertaining to banks, credit cards, personal details, and government policies are transferred from place to place with the help of networking infrastructure. The high connectivity of the World Wide Web (WWW) has left the world "open". Such openness has resulted in various networks being subjected to multifarious attacks from vastly disparate sources, many of which are anonymous and yet to be discovered. This growth of the WWW coupled with progress in the fields of e-commerce and the like has made the security issue even more important.

A typical method for security that is used to prevent data from falling into wrong hands is encryption. Some encryption techniques like RSA ([13]) which use asymmetric keys, involve algebraic multiplications with very large numbers. The cost that has to be paid in implementing encryption in networks is high owing to this computational complexity. While other techniques like the DES ([12]) which use symmetric keys are less secure computationally than their asymmetric counterparts. Given the amount of computing power that is available, and considering also the growth of distributed computing, it is possible to break into the security offered by many such existing algorithms. So, any alternative to encryption is welcome so long as the level of security is the same or higher. Also, such an alternative should be more efficient in its usage of resources.

In practice, in a computer network, data is transferred across nodes in the form of packets of fixed size. Any form of security required is obtained by implementing cryptographic algorithms at the application level on the data as a whole. Then, the enciphered data is packetized at lower levels (in the OSI model) and sent. Any intruder able to obtain all the packets can then obtain the enciphered data by appropriately ordering the data content of each of these packets. Then, efforts can be made to break the cryptographic algorithm used by the sender. In the process of transmission, if it is possible to prevent any information release as to the structure of the data within the packets, an intruder would know neither the nature of the data being transferred nor the ordering of the content from different packets. This is what our algorithm achieves by using a one-time pad like scheme at the source.

The essential idea in our algorithm is to break the data that is to be transferred into many chunks, which we call "parts". These parts when put together form the whole data but only if done so in a particular way, just like in a "jigsaw" puzzle. The only way of doing so is known to the receiver for whom the data is intended. Any unauthorized node does not have enough information to carry out the right method of obtaining the parts from the packets and joining them, and then *knowing* it is correct (which is the property of the one-time pad). We have incorporated efficient techniques that enhance the security of the scheme, and at the same time implement the desired features. The concept on which our algorithm hinges heavily is that of the one-time pad. It was first proposed by Vernam in [1]. A formal proof of the perfect-secrecy property of the one-time pad was later provided by Shannon in [2]. As can be seen from the reference, the one-time pad offers unconditional security as against conditional or computational security of all encryption algorithms currently known.

## II. Related Work

Owing to several issues mostly pertaining to key management, the theoretical one-time pad has been tough to implement practically. Numerous attempts have been made but under varying assumptions and conditions. One of the most recent has been [4] where one-time pads are used to protect credit card usage on the Internet. In [5], it has been argued that unconditional security can be obtained in practice using non-information -theoretically secure methods. This approach maintains that in the practical world, nobody can obtain complete information about a system owing to real-world parameters like noise. Likewise, [6], [7] and [8] provide implementations of one-time pads and unconditional security but under assumptions about the environment and/or adversary. [9] provides a quantum cryptographic view of one-time pads.Technological and practical limitations constrict the scalability of such methods! . Chaotic maps are used to generate random numbers in [10], and these are used to build symmetric encryption schemes including one-time pads. Chaos theoretical methods though providing non-traditional methods of random number generation, are prone to cryptanalysis owing to the still-existent pseudorandom nature, and impose numerous restrictions on data size as is the case with [10].

In this paper, we provide an efficient implementation of the one-time pad without making assumptions or imposing restrictions, the likes of which are true of the references quoted in the previous paragraph. In the process, the core issues including key management are addressed and dealt with effectively. Due to its general nature, our algorithm can be deployed in most real-life networks without a fundamental change in the idea.

## III. The Algorithm

In this algorithm, we use the concept of Message Authentication Code (MAC) as suggested in [3] to authenticate messages. For a packet of data, the MAC is calculated as a function of the data contents, the packet sequence number and a secret key known only to the sender and the receiver, and then it is appended to the packet. On receiving a packet, the receiver first computes the MAC using the appropriate parameters, and then performs a check with the MAC attached to the packet. If there is no match, then the receiver knows that the packet has been tampered with. For a detailed analysis, the reader is referred to [3].

Let the size of a packet in a network be denoted as PS. PS has a value of 1024-bits or 4096-bits typically. The prerequisite of the algorithm is the knowledge for the sender and the receiver only of a number P, exchanged a priori, of size $k*PS$, where $k$ is a natural number. We can consider P in terms of blocks of size PS each, as $P_1, P_2, \cdots, P_k$. Thus P is the number obtained by the concatenation of the $P_i$ blocks for $i$ from 1 to $k$, that is, P is $P_1 P_2 \cdots P_k$.

In this algorithm, we only consider *k-1* parts of a data at a time, where each part is of any size less than or equal PS-2 (a detailed analysis of the algorithm is presented in the next section). If the entire data is not covered in these *k-1* parts, then the algorithm can be repeated by considering the next *k-1* parts and so on. Also, a random number of size PS is required to be generated for every *k-1* blocks processed. Denote this random number as R.

The steps at the sender's end of the algorithm are as follows:

*S(D)*

1) "Tear the data D into N parts arbitrarily". Consider the first *k-1* parts of D. Call them

$D_1, D_2, \cdots, D_{k-1}$. Prefix and suffix each part by the binary digit '1'.
2) Perform the operation $1D_i1 \oplus P_i \, \forall \, i$ from 1 to *k-1*. Denote them as $D'_1, D'_2, \cdots, D'_{k-1}$.
3) Form $D'_k$ as $D'_k = R \oplus P_k$.
4) Perform *transform(P,R)*.
5) Generate a new random number and assign it to R.
6) Repeat steps 1, 2, 3, 4, 5 for the next *k-1* blocks of data, and so on until all N parts are processed.

Now the packets actually transferred are formed from the $D'_i$ blocks, packet sequence number and the MAC (calculated as described earlier).

At the receiver's end, the steps are as follows:

*R(D')*

1) Perform a check on the MAC for each packet. If satisfied, GOTO next step.
2) Order packets according to the packet sequence number.
3) For each group of *k* packets perform the following:
   - Perform $P_i \oplus D'_i \, \forall \, i$ from 1 to *k-1*
   - Perform $P_k \oplus D'_k$ and obtain R.
   - Remove the leading and trailing '1' of all the values obtained from the previous two steps.
   - Perform *transform(P,R)*.
   - *continue*

The algorithm for the operation *transform(P,R)* is as follows:

*transform(P,R)*

1) Set $P_i \leftarrow P_i \oplus R \, \forall \, i$ from 1 to *k-1*.
2) Set $P_k \leftarrow P_k * R$.

For the transfer of the next data, the following is done. The sender knows the number of parts of the previous data and hence knows the value of N % *k*, where % denotes the 'modulo' operation. Now, this value subtracted from *k* provides the $P_i$s unused in the last run of the loop in the algorithm. The next data to be transferred is first broken into parts as before. For the first run of the loop in the algorithm, the first *k-1-(N%k)* (here, this is the old value of N) parts are only considered and the run executed. For the remaining runs, we consider *k-1* parts as before, and the algorithm is continued. This is done for all subsequent data transfers between the sender and the same receiver.

## IV. AN ANALYSIS

### A. Discussion

The essential idea in the algorithm is to split the data into pieces of arbitrary size (rather, the size is not arbitrary since it is bound by 0 and PS-2 but is allowed to take any value between the limits), transmit the pieces securely, and provide a mechanism to unite the pieces at the receiver's end. Towards this, the role of the number P is to provide a structure for the creation of the "jigsaw" puzzle. This structure masks the pieces as well as protects the data within. The structure is changed periodically with the knowledge of both the sender and the receiver to prevent an adversary gaining information about it.

The $\oplus$ function is reversible and can be easily performed by the receiver since he knows the secret number *P*. However, an adversary without knowing *P* cannot obtain any additional information. This is because of the following.

Given a ciphertext C of length PS, and a random secret key *P* of length PS, the probability of any particular key is the same. If it is possible to guess the message M such that $C = M \oplus P$, then it is possible to determine the value of *P*. Since every secret key *P* is equally likely, there is no way of guessing which of the possible messages of length PS or less was sent. In other words, let us assume that the adversary has as much knowledge of the cryptographic mechanism as the receiver does (except, of course, knowledge of the secret key *P*). Now, the only knowledge that an adversary has about both M and *P* is that one is a binary string of length PS while the other is of length less than PS. Then, knowledge of the ciphertext C does not provide the adversary with any more information pertaining to either M or *P*. This is the information-theoretic security property of a one-time pad (for more details, refer [2]). ! The value $D'_i$ obeys this one-time pad property since both $P_i$ and $D_i$ are random numbers as far as the adversary is concerned. Thus, it is impossible for an adversary to get any more information given this value. Thus, all of $D_i$ remain completely unknown to the adversary.

Now, after *k* parts are processed, it is not possible to repeat the values $P_1, P_2, \cdots$. Else, by manipulations due to the reversible nature of $\oplus$, some information might be leaked to the adversary. Therefore, the next set of values have to be changed, and towards this a random number of equal size is used. Also, this random number is to be conveyed to the receiver without any adversary knowing its value. Hence, we introduce the random number as the $k$th part. This random number is used to calculate the next P to be used. Since the initial P was a secret for the adversary, and so is the random number, the new value of P is also a secret. Thus, the security of the data transmission is ensured. In our model of the one-time pad, data is of effective size less than ($k-1$)*PS bits and the key 'xor'ed with the data at each stage is the number $P_1 P_2 \cdots P_{k-1}$. The random number R is used as an input to a function (namely *tran! sform()*) to generate the key for the next run. The value $P_k$ is used to securely convey the random number R, that is generated by the sender, to the receiver.

The need for prefixing and suffixing every part with the bit '1' is explained below. By 'xor'ing a part with a random (with respect to the adversary) key of size PS, we are effectively embedding the data in some place in the key. However, there is a problem of the receiver not knowing where the data is embedded. This is best illustrated with an hypothetical example. Suppose the part is the binary sequence *01101*, and the key is the sequence *11000110* where PS=8. Now embedding the data at position 3 from the left in the key, we get the value as *11011000*. When the receiver performs the 'xor' of this value with the key, the value he obtains is *00011010*. Now, there is no way of knowing whether the leading or the trailing zeros are part of the data or not. This is the reason for the use of the single bit '1' to prefix and suffix each part. As an illustration of this, we continue with the same example. Note that the part is necessarily smaller than the key by at least 2. Here the part under question is affixed and prefixed by '1', and then embedded in the key at position 2, say (position 3 can no longer be used). In this case, the resultant value is *10011101*. On performing the 'xor' with the key, the receiver obtains the value *01011011*. From this value, he considers only the binary sequence sandwiched between the first and last occurrences of the bit '1' which is the intended part.

A point of worry is that the arbitrary "tearing" of data might result in a huge expansion of data. That is, due to the randomness in the splitting, the amount of data sent to convey some fixed amount of information might be huge. To avoid this, the algorithm provides the flexibility to fix a lower limit on the size a part can take. This ensures that a minimum amount of information is transferred to the receiver by each part. Thus, the overheads in the algorithm can be suitably controlled by the lower-limit size. It has been proved by Shannon in [2] that in a one-time pad, as long as the data is of a size lesser than or equal to the key used, then the perfect-secrecy property is maintained. This justifies the statements made in this paragraph. In fact, for least overhead, each part could be made of size PS (in this case, there is no need to prefix and suffix with bit '1'). Our algorithm is however a general case of this, providing more security to lessen the probability of a serendipitically successful attack.

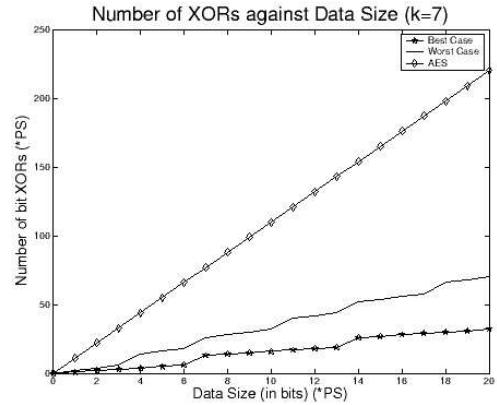

Fig. 1. Comparison based on XORs alone

The security offered by the algorithm is the same as that provided by one-time pad - information-theoretic security. This is evident from the fact that the jigsaw is structured by an embedding of the data in the key using the 'xor' operation. However, unlike the one-time pad, keys used in our algorithm are not completely uncorrelated since the next key is formed as a function of the current key and a random value. Thus, under conditions of information leakage, the security offered by our approach fails while the one-time pad continues to offer the same level of security to the unexposed data. The nature of security under such conditions has not been studied here.

For private-key message authentication purposes, another secret key might be needed. However, since any PS-size block of P cannot be guessed from its first use, it is possible to use any of them or parts of it (in case a smaller key is desired) for message authentication by the calculation of MAC. This information is also to be exchanged a priori along with the number P as well as the value of $k$. For different sender-receiver pairs, different secret numbers P are needed. It is necessary to take care to see that this indeed is the case.

*B. Implementation Issues*

In our algorithm, for the transfer of N parts of data, there are N additions performed. Apart from these operations, the value of P is changed $\lfloor \frac{N}{k} \rfloor$ times. At each change, there are *k-1* additions and 1 multiplication. Hence, in total our algorithm requires $N + \lfloor \frac{N}{k} \rfloor * (k-1)$ additions and $\lfloor \frac{N}{k} \rfloor$ multiplications on PS-bit blocks for the transfer of N parts of data. There is a degree of uncertainty involved in the amount of information transferred in a part as the splits happen arbitrarily. This uncertainty can, however, be removed marginally by imposing a lower limit on the size a part can take thus ensuring a transfer of minimum information in each part. Let the value of *k*=7. To transfer data of size 10*PS, assuming the best case of PS- size parts, we have N=10, and therefore we effectively require 16*PS bit additions and 1 PS-bit multiplication. In the worst case, assuming the lower limit ! on size of each part to be $\frac{PS}{2}$ , we require 20 parts to transfer the same amount of data in which case we would effectively require 32*PS bit additions and 2 PS-bit multiplications. A graph of the best case(corresponding to each part being of maximum size = PS) and the worst case (corresponding to each part being of minimum size = $\frac{PS}{2}$) in the number of additions versus data size is shown in figure 1. Figure 2 depicts the plot of the number of additions in the best and worst cases as a function of the value *k* (assuming N=10). Figure 3 plots the number of multiplications as a function of *k* (again assuming N=10). It is important to note that as *k* increases, the size of P also increases. Therefore, an optimum value of *k* should be arrived at by analysing these graphs and statistics.

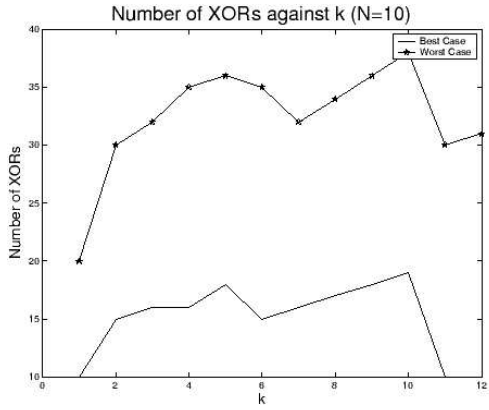

Fig. 2. Change in number of XORs with *k*

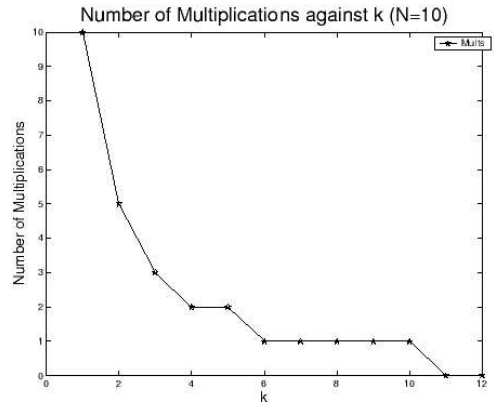

Fig. 3. Change in number of multiplications with *k*

A comparison with other encryption algorithms is valid only when the application of those algorithms is for secure data transmission. From this viewpoint, the statistics presented above compare favourably with respect to encryption algorithms like Advanced Encryption Standard(AES) [11]. In AES, for input block size of 128-bits and key length of 128-bits, there are atleast 11 'xor' operations apart from matrix multiplications, table lookups and vector shifts. When AES is scaled to an input of 10 PS-size blocks, it requires 110*PS bit additions, far more operations than our algorithm functioning in the worst case as can also be witnessed in the figure 1 (we have not included the matrix multiplications, table lookups and vector shifts in our calculations). Also, our algorithm does not transform the data except for the 'xor' operation. This operation and its inverse can be easily computed. Hence, as compared to encryption algorithms like AES, DES ([12]) and RSA ([13]), the data processing time is least for our algorithm.

Our algorithm lends itself to parallelism in implementation in software as well as dedicated hardware. The execution of the operation *transform(P,R)* should follow the processing of *k* blocks. This sequentiality cannot be avoided. However, the processing of the blocks can be done in a parallel manner. In principle, the algorithm can be implemented efficiently using *k* 'xor' gates, as shown in figure 4. Here, the first three cycles of 'xor's are depicted with the respective inputs and outputs. If number of gates is also a constraint, for best performance, the value of *k* should be decided accordingly.

## V. CONCLUSION

In this paper, we adopt a "jigsaw" approach to secure data transfer in networks. The data to be sent is broken into parts of arbitrary sizes. Enough information is provided efficiently and securely to enable the receiver to solve the "jigsaw" puzzle. The transfer of the parts is done securely without leaking *any* information to the adversary regarding the data. We have illustrated a method of implementing message authentication by private key without the exchange of any more information. The concept of the one-time pad is implemented in the course of the algorithm resulting in information-theoretic security of data transfer. The issue of key management is addressed by firstly the exchange of a large number a priori, and then subsequent modifications to the large number at regular intervals. These modifications are designed such that their outputs seem random to the adversary. Also, flexibility in the form of a means of control is provided in the algorithm to monitor and check the overhead resulting because of the data expansion due to the arbitrary splitting.

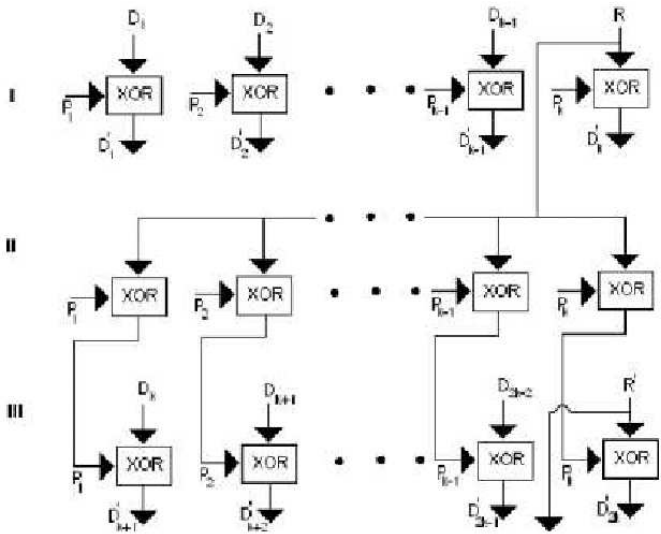

Fig. 4. Parallelism in algorithm (R' denotes a new random number)